\documentclass[prd,aps,floats,twocolumn,floatfix]{revtex4}
\usepackage{graphicx, epsfig}
\newcommand{\be}{\begin{equation}}
\newcommand{\ee}{\end{equation}}
\newcommand{\bea}{\begin{eqnarray}}
\newcommand{\eea}{\end{eqnarray}}

\begin{document}
\setlength{\unitlength}{1mm}
\title{Constraining the shape of the CMB: a Peak-by-Peak analysis.}
\author{Carolina J. \"Odman$^\sharp$, Alessandro Melchiorri$^\flat$,Michael P. Hobson$^\sharp$ and Anthony N. Lasenby$^\sharp$}
\affiliation{ $^\sharp$ Astrophysics Group, Cavendish Laboratory, Cambridge University, Cambridge, U.K.\\$^\flat$ Astrophysics, Denys Wilkinson Building, University of Oxford, Keble road, OX1 3RH, Oxford, UK}
\begin{abstract}The recent measurements of the power spectrum of Cosmic Microwave Background
anisotropies are consistent with the simplest inflationary scenario 
and big bang nucleosynthesis constraints.
However, these results rely on the assumption of a 
class of models based on primordial adiabatic perturbations, cold dark matter
and a cosmological constant.
In this paper we investigate the need for deviations from
the $\Lambda$-CDM scenario by first characterizing the spectrum
using a phenomenological function in a $15$ dimensional 
parameter space. 
Using a Monte Carlo Markov chain approach to 
Bayesian inference and a low curvature model template we then 
check for the presence of new physics and/or systematics
in the CMB data.
We find an almost perfect consistency between the phenomenological 
fits and the standard $\Lambda$-CDM models.
The curvature of the secondary peaks is weakly constrained 
by the present data, but they are well located. 
The improved spectral resolution expected from future satellite 
experiments is warranted for a definitive test of the scenario. 
\end{abstract}
%\bigskip]
%\pacs{PACS Numbers: }
\maketitle
\section{Introduction.}
The recent observations of the
 cosmic microwave background (CMB) anisotropies power spectrum 
(\cite{toco},\cite{b97}, 
\cite{Netterfield},\cite{halverson},\cite{lee}, 
\cite{cbi}, \cite{vsa}, \cite{archeops}, \cite{acbar}, 
\cite{ruhl},\cite{vsae})
have presented cosmologists with the possibility of studying
the large scale properties of our universe with unprecedented 
precision. As is well known (see e.g. \cite{review}),
the structure of the theoretical CMB spectrum, given mainly by the relative
positions and amplitude of the so-called acoustic peaks, is sensitive
to several cosmological parameters.
The existing CMB data sets are therefore being analyzed 
with increasing sophistication (see \cite{koso} and \cite{sko} for
important advancements) in an
attempt to measure the undetermined 
cosmological quantities.
The most recent analyses of this kind (\cite{debe2001}, 
\cite{pryke}, \cite{stompor}, \cite{wang}, \cite{cbit}, 
\cite{vsat},\cite{bean},\cite{saralewis},\cite{mesilk}, \cite{archeops2},
\cite{slosar},\cite{wang2}) have revealed an 
outstanding agreement between the data and the
inflationary predictions of a flat universe and of
a primordial scale invariant spectrum of adiabatic density
perturbations.
Furthermore, the CMB constraint on
the amount of matter density in baryons $\omega_b$ is 
now in very good agreement with the independent 
constraints from standard big bang nucleosynthesis (BBN)
 obtained from primordial deuterium (see e.g. \cite{burles}, 
\cite{hansen}) and consistent within $2$-$\sigma$ with
those derived from the combined analysis of $^4He$ and $^7Li$ 
(\cite{cyburt}). 
Finally, the detection of power around the expected third
peak, on arc-minutes scales, provides a new and independent 
evidence for the presence of non-baryonic dark matter
(\cite{mesilk}).

The data therefore suggests that our present
cosmological model represents a beautiful and elegant 
theory able to explain most of the observations.

However, the CMB result relies on the assumption of 
a particular class of models, based on
adiabatic, {\it passive} and {\it coherent} 
(see \cite{andy}) primordial fluctuations, and cold dark matter.
In the following we refer to this class of models as
$\Lambda$-Cold Dark Matter ($\Lambda$-CDM).

This weak point, shared by most of the current studies, 
should not be overlooked: it has been recently shown,
for example, that the very legitimate inclusion of gravity waves 
(see e.g. \cite{efstathiou}, \cite{gw}) 
or isocurvature modes (\cite{kxm}, \cite{trotta}, 
\cite{amendola}) 
into the analysis can completely erase most of the constraints
derived from CMB alone.

Furthermore, since even more exotic modifications like 
quintessence (\cite{caldwell}), topological 
defects (\cite{bouchet},\cite{dkm}), 
broken primordial scale invariances (\cite{alexandra},
\cite{bend}, \cite{covi}), 
extra dimensions (\cite{bisilk}) or unknown systematics 
(just to name a few) can be in principle 
considered, one should be extremely cautious in making any 
definitive conclusion from the present CMB observations.

It is therefore timely to investigate if the present CMB 
data are in complete agreement with the $\Lambda$-CDM scenario
or if we are losing relevant scientific informations 
by restricting the current analysis to a subset
of models (see e.g. \cite{tegza}).

In the present {\it paper} we check to what extent 
modifications to the standard $\Lambda$-CDM scenario
are {\it needed} by current CMB observations with two 
complementary approaches:
First, we provide a model-independent analysis by
fitting the data with a phenomenological function
 and characterizing the observed multiple peaks. Phenomenological 
fits have been extensively used in the past and recent CMB 
analyses (\cite{rocha},
\cite{page}, \cite{miller2k2}, \cite{podariu}, 
\cite{boghdan}, \cite{douspis}).
Our analysis differs in two ways: we include the latest 
CMB data from the Boomerang (\cite{ruhl}), VSAE (\cite{vsae}), 
ACBAR (\cite{acbar}), and Archeops (\cite{archeops}) experiments and
we make use of a Monte Carlo Markov Chain (MCMC) algorithm, 
which allows us to investigate a large number of parameter 
simultaneously ($15$ in our case). 

We then compare the position, relative amplitude and width of the
peaks with the same features expected in a $4$-parameters 
model template of $\Lambda$-CDM spectra. 
By doing a peak-by-peak comparison between the theory and the
phenomenological fit which is based on a much wider set 
of parameters, we then verify in a systematic way the agreement 
 with the standard theoretical expectations.

As a by-product of the analysis, we present a set of 
cosmological diagrams that directly translate, 
under the assumption of $\Lambda$-CDM, 
the constraints on the features in the spectrum into 
bounds on several cosmological parameters.
These diagrams offer the opportunity of quick,
by-eye, data to model comparison.

Our paper is organized as follows: In section II we 
discuss the phenomenological representation of the power
spectrum, the analysis method we used and the MCMC algorithm. 
In section III we present our results. Finally, in section IV, 
we discuss our conclusions.

\medskip
\section{Phenomenological representation.}
\medskip

We model the multiple peaks in the CMB angular spectrum 
by the following function:

\begin{equation}
\ell(\ell+1)C_\ell/2\pi = \sum_{i=1}^N \Delta T_i^2 \exp 
(-(\ell-\ell_i)^2/2\sigma_i^2)
\end{equation}
where, in our case, $N=5$.
In order to avoid degeneracy of overlapping gaussians, we 
parametrise the centers of the secondary gaussians 
as functions of the positions of the previous gaussians:
\begin{eqnarray}
l_2 & = & l_1 ( 1 + \alpha) \\ \nonumber
l_3 & = & l_1 ( 1 + \alpha + \beta) \\ \nonumber
l_4 & = & l_1 ( 1 + \alpha + \beta + \gamma) \\ \nonumber
l_5 & = & l_1 ( 1 + \alpha + \beta + \gamma + \delta) \\ \nonumber
\end{eqnarray}

We use this formula to make a phenomenological fit
to the current CMB data, constraining the
values of the $15$ parameters $\Delta T_i$, $\ell_i$ and 
$\sigma_i$.

The use of gaussian-shaped function to describe the
CMB spectrum is now becoming a standard method 
in the literature (see e.g. \cite{page}, \cite{boghdan}, 
\cite{debe2001}, \cite{cbit}). 
A major difference with respect to previous works is that we 
are using only one fitting function, 
varying all its parameters simultaneously, 
while in general peaks are characterized with one
single function in different selected regions of the spectrum, in
correspondence with the expected peaks. 

The advantage of a single fitting function is a better control 
of the correlations between the phenomenological 
parameters as we show in the next section where we report
the values of the correlation matrix.

Recently, Douspis and Ferreira (\cite{douspis})
used a Gaussian plus an oscillating function as a phenomenological model.
The method used here is more general, in the sense 
that we allow independent amplitudes and widths of 
the secondary peaks as well as impose no periodicity.

We use the CMB data as listed in table \ref{cmbdata}, spanning 
the range $50 \le \ell \le 2500$.

For all the experiments we use the publicly available window 
functions and correlations in order to compute the
expected theoretical signal $C_B$ inside the bin.
The likelihood for a given phenomenological model is defined by 
 $-2{\rm ln} {\cal L}=(C_B^{ph}-C_B^{ex})M_{BB'}(C_{B'}^{ph}-C_{B'}^{ex})$ 
where  $M_{BB'}$ is the Gaussian curvature of the likelihood  
matrix at the peak. When available, we use the 
lognormal approximation to the band-powers.

We marginalize over the reported Gaussian distributed  
calibration error for each experiment and we include 
the beam uncertainties by the analytical marginalization 
method presented in (\cite{sara}).

\begin{figure}[ht]
\centerline{\rotatebox{270}{\psfig{figure=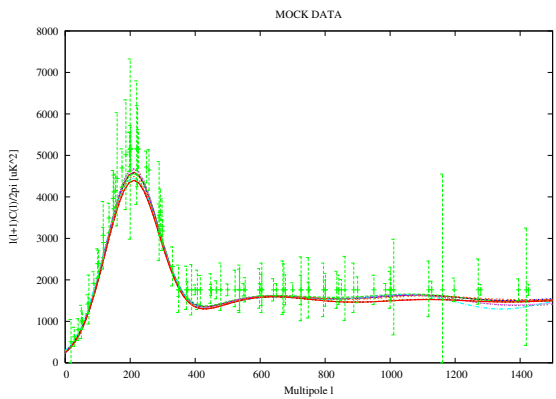,width=5.3cm}}}
\caption{The 10 best MCMC samples fitting mock data. The data were generated by convolving the experimental window functions with a power spectrum consisting of a first peak followed by a flat line.}
\label{flatfit}
\end{figure}

We perform our analysis on $3$ different data sets: 
the full data set, the low-frequency (LF) data (experiments that
covered frequencies in the electromagnetic spectrum of
$90$ GHz and below) and a high-frequency 
(HF) data (frequencies higher than $90$ GHz).
The reasons of this choice are twofold: First, any discrepancy 
between the 2 analyses would hint towards the presence of undetermined foreground. Secondly, this facilitates the comparison with future and soon to be released observed power spectra at ``low'' frequencies, like those expected from the MAP satellite. We test the stability of our result by including a set of older CMB experiments as reported in table \ref{cmbdata} and we also check that our model contains no bias towards the presence of peaks by fitting a set of mock data from a spectrum (see fig. \ref{flatfit}).

\begin{table*}\caption{\label{cmbdata}List of experimental data used in this study. We examine low and high frequency experiments together and separately. We then check our results by including some older experiments.}
\begin{center}\begin{tabular}{lcc}
\hline
Experiment & $l$ range & reference \\\hline\hline
{\bf High Frequency Experiments (HF)}:\\\hline
Acbar  & $150$ -- $3000$& Kuo  {\it et al.}, 2002\\\hline
Archeops  & $15$ -- $350$& Benoit  {\it et al.}, 2002\\\hline
Boomerang 98  & $50$ -- $1000$& Ruhl  {\it et al.}, 2002\\\hline
Maxima  & $73$ -- $1161$ & Lee et al., 2001, Ap. J. 561 (2001) L1-L6\\\hline
{\bf Low Frequency Experiments (LF)}:\\\hline
DASI  & $117$ -- $836$ & Halverson {\it et al.}, Ap. J, 568, 38, 2002\\\hline
CBI   & $400$ -- $1450$ & Pearson {\it et al.}, 2002\\\hline
VSAE  & $160$ -- $1400$ & Grainge {\it et al.}, 2002 \\\hline
{\bf Older Experiments}:\\\hline 
TOCO 97  & $63$ -- $194$ & Miller {\it et al.}Ap. J. Supp., 140, 115, 2002\\\hline
TOCO 98  & $128$ -- $409$ & Miller {\it et al.}Ap. J. Supp., 140, 115, 2002\\\hline
MSAM & $90$ -- $400$ & Wilson, et al., 2002\\\hline
QMASK & $105$ -- $359$ &  Xu {\it et al.}, Phys. Rev. D65, 2002\\\hline
Python V  & $67$ -- $267$ & Coble {\it et al.}, 2001 \\\hline
Boomerang NA  & $50$ -- $400$ & P. Mauskopf {\it et al.}, Ap. J. Letters, 536, L59, 2000\\\hline
\end{tabular}
\end{center}
\end{table*}

\begin{figure}[ht]
\centerline{\rotatebox{270}{\psfig{figure=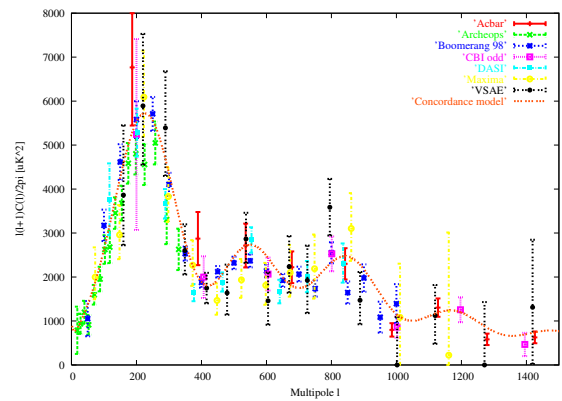,width=5.3cm}}}
\caption{The CMB data used in this analysis: 
beam and calibration errors are not included.}\label{ct}
\end{figure}

The phenomenological fit is operated through an MCMC algorithm.
The MCMC approach is to generate a random walk through parameter space that 
converges towards the most likely value of the parameters and samples the 
parameter space following the posterior probability distribution. 
In the general case, the number of parameters and their priors have to be 
defined. 
There is no limit in resolution (except numerical precision of the computer).
The priors define the volume in parameter space in which the random walk takes place.\\
At each iteration, 
a point $\mathbf{x}_{n+1}$ is randomly selected in 
the $m$-dimensional parameter space. Its likelihood is evaluated by 
comparing to the data. A sample is said to be accepted into the Markov chain 
or not, depending on the following acceptance criterion:
\begin{equation}
u \leq \min \left\{ 1, \frac{p(\mathbf{x}_{n+1}) 
\mathcal{L}(\mathbf{x}_{n+1})}{p(\mathbf{x}_{n}) 
\mathcal{L}(\mathbf{x}_{n})}\right\},
\nonumber
\end {equation}
where $u$ is a random number sampled from a uniform distribution on the interval $[0,1]$, $p(\mathbf{x})$ is the prior and $\mathcal{L}(\mathbf{x})$ 
is the likelihood, containing the information from the data.
At the end of the MCMC routine, the samples are counted and their number 
density projected onto one or more dimensions 
is proportional to the marginalised posterior distribution of the parameters. 
If the posterior follows a gaussian distribution, the best fit value is obtained by  averaging over the samples. The MCMC procedure is described in more detail
in \cite{saralewis} and \cite{christensen}.\\

We use wide uniform priors for each parameter, in which our resulting
likelihoods are fully encompassed, except for the widths of the secondary gaussians which are not well constrained by the data.
We run the MCMC routine in order to get $\sim 7000$ samples for each subset of data. Also, in order to check that out parametrisation is not biased towards 
the presence of peaks, we test it to fit a flat spectrum. 
We generate a set of mock data by convolving a power spectrum consisting of a peak and a flat line with the experimental window functions. 
The results show that a flat power spectrum is easily 
recovered within our priors, as shown in figure \ref{flatfit}.

We then consider a flat, adiabatic, 
$\Lambda$-CDM model template of CMB angular power
spectra, computed with CMBFAST (\cite{sz}), sampling the various
parameters as follows:
$\Omega_{cdm}h^2\equiv \omega_{cdm}= 0.01,...0.40$, in steps of  $0.01$;  
$\Omega_{b}h^2\equiv\omega_{b} = 0.001, ...,0.040$, 
in steps of  $0.001$ and $\Omega_{\Lambda}=0.0, ..., 0.95$, 
in steps of  $0.05$. 
The value of the Hubble constant is not an independent 
parameter, since: 
 \begin{equation} 
h=\sqrt{{\omega_{cdm}+\omega_b} \over {1-\Omega_{\Lambda}}}. 
\end{equation}
We vary the spectral index of the primordial density perturbations 
within the range $n_s=0.60, ..., 1.40$ (in steps of  $0.02$).

For each model in the template we then consider the 
corresponding values $\Delta T_i$, $\ell_i$, $\sigma_i$ such that
the formula in Eq. (1) represents the best fit to its shape. Indeed, the $\Delta T_i$ do not exactly correspond to the amplitudes of the peaks, as our spectrum is a sum of gaussians and the power from each gaussian contribute over the whole spectrum.
We find that, restricting the range in $\ell$ to
$50,...,1500$, equation (1) approximates the shape of 
the spectra in our template well (better than $\sim 10 \%$ in $C_{\ell}$).

We also check that the use of different phenomenological functions
such as lorentzians or log-normals has no relevant effect on our results.

It is important to note that we restrict the parametrisation of
our template of theoretical models to a set of only $4$ parameters.
However, because of the 'cosmic degeneracy' in the CMB observables, 
this is enough to describe the possible
shapes of the CMB spectra in the $\Lambda$-CDM scenario.
Increasing the optical depth $\tau_c$ or adding a background of gravity
waves, for example, is nearly equivalent to changing
some of the parameters already considered like $n_S$ and $\omega_b$.
On the other hand we characterize the peaks in the spectrum
with a phenomenological fit based on $15$ parameters, which
allows independent positions, amplitudes and widths
of the observed features.

The comparison between the model-independent values
$\Delta T_i$, $\ell_i$ and $\sigma_i$ obtained by fitting the data
with Eq. (1) and the corresponding values expected in 
the template of theoretical models represents therefore
a strong check of the theory and can give hints for the
presence of systematics and/or new physics.

\section{Results}
In Table \ref{fmaev} we report the $68 \%$ limits on $\Delta T_i$, $\ell_i$ and
$\sigma_i$ of Eq. 1 obtained by analyzing the present CMB data
with an MCMC procedure for each subset of CMB data.
We also report the constraints on 
combinations of those parameters that are more
directly connected to the cosmological parameters
(see the discussion below).

In Table \ref{corre} we report the correlation matrix between the parameters
of our phenomenological fit. As we can see, important correlations
exist between the parameters: for example, the amplitude
of the peaks is highly anti-correlated with  the widht of the adiacent 
gaussians. This further illustrate the utility of analyzing the data
with a single fitting function in order to properly evaluate 
the statistical significance of the oscillations.

\begin{figure}[ht]
\centerline{{\psfig{figure=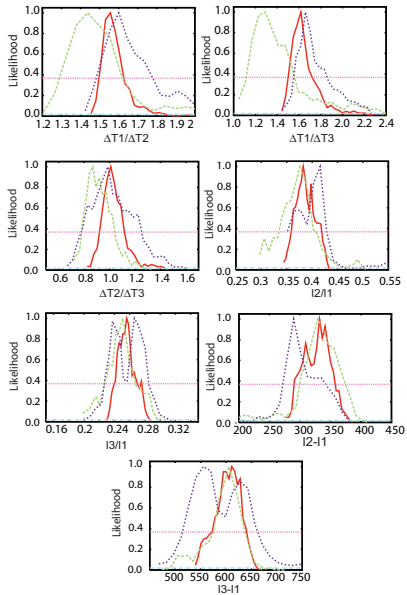,width=5.3cm}}}
\caption{Marginalised likelihood functions for CMB observables.
The red (solid), green (dashed) and blue (dotted) curves correspond
to the following three data sets, respectively: Full data set, LF, HF.}
\label{histograms}
\end{figure}

\begin{figure}[ht]
\centerline{\rotatebox{270}{\psfig{figure=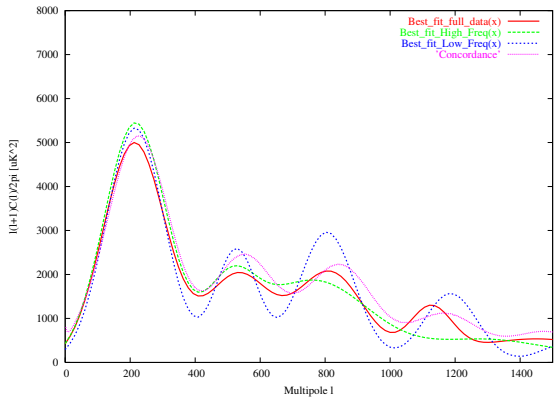,width=5.3cm}}}
\caption{Comparison of the concordance cosmological power spectrum and the best-fit phenomenological power spectrum for each data subset.).
}
\label{bestfits}
\end{figure}

Figure \ref{histograms} shows the marginalised likelihood functions
of the important CMB observables, and the agreement between the three
data subsets. Figure \ref{bestfits} shows the best-fit
cosmological and phenomenological models. Although our sums
of gaussians comprises any cosmological power spectrum to
within $10 \%$, their preferred shape differs from the cosmological
model owing to the large parameter space allowing to fit the data
very well.

\begin{figure}[ht]
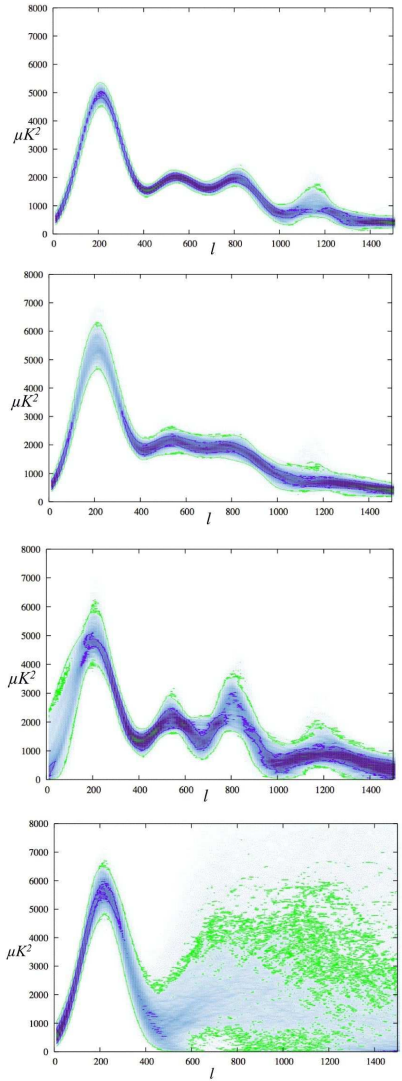

%\centerline{{\psfig{figure=rnbi.tsunami.ps,width=5cm}}}
%\centerline{{\psfig{figure=rnb.tsunami.ps,width=5cm}}}
%\centerline{{\psfig{figure=rni.tsunami.ps,width=5cm}}}
%\centerline{{\psfig{figure=o.tsunami.ps,width=5cm}}}
\centerline{{\psfig{figure=rnbi.epsf,width=5.3cm}}}
\centerline{{\psfig{figure=rnb.epsf,width=5.3cm}}}
\centerline{{\psfig{figure=rni.epsf,width=5.3cm}}}
\centerline{{\psfig{figure=o.epsf,width=5.3cm}}}
\caption{$68 \%$ and $95 \%$ confidence levels in the $\Delta T^2$,
$\ell$ plane. The first three panels correspond to the full data set, HF data and LF data, respectively. The last panel shows the same result obtained by analysing the old data only. It is found to be consistent with the more recent data, and provides no constraints above $\ell \sim 400$ where the data provide no measurements. This shows where the power is well constrained by the data using our phenomenological function. Where no limits of power are found within $68\%$, it means that the data are not sufficient to constrain the power at that angular scale.}
\label{tsunami}
\end{figure}

Figure \ref{tsunami} shows $68\%$ and $95\%$ confidence levels in power in the $\ell$-$\Delta T^2$ plane. This shows the most likely path of the power spectrum. These figures are generated as follows: First, we remind the reader that the number density of MCMC samples is proportional to the marginalised posterior. We divide the $\ell$-$\Delta T^2$ plane into pixels of size $10 \times 10$. For each pixel, we count the phenomenological spectra that go through it. Hence, this is a density plot of our MCMC samples.
The absence of a $68\%$ region at certain $\ell$ ranges indicates that more data are needed to constrain the power.

From figure \ref{tsunami}, it appears that LF experiments provide constraints at high multipoles better than HF experiments which constrain the power at intermediate scales strongly. HF experiments also provide tight constraints at low $\ell$, mainly due to Archeops.

We also show the same result obtained by using the old data {\it only}. We find that the analysis provides no constraints above $\ell \sim 400$, as expected from the data. It is also consistent with the Archeops observations. Hence, when combining the old data with the LF data, HF data or both, our results remain essentially the same.

The values are in reasonable agreement with the results
obtained by similar analyses (see e.g. \cite{debe2001}, 
\cite{boghdan}, \cite{douspis}, \cite{cbit}) and point 
towards the presence of multiple peaks in the spectrum.

The low frequency and high frequency experiments yield
consistent results, showing that possible systematics due
to galactic foregrounds are under control.
However it is worth noticing that the LF data
are more consistent with higher amplitude of secondary peaks.
These experiments also use different techniques. The HF experiments are  bolometers whereas the LF experiments are interferometers. The different nature of experimental uncertainties as well as their evaluation might suffer from different weaknesses contributing to enhance this contrast.

The CBI and ACBAR data at high $\ell$ are in agreement with
the expected damping tail (see e.g. \cite{tail}). However,
the poor spectral resolution ($\Delta{\ell}\sim 200$) does not allow
us to constrain subsequent peaks. 

Still, it is interesting to compare the values obtained 
with those expected in the $\Lambda$-CDM scenario with different 
priors, as we do in the last three columns of
Table \ref{fmaev} (see caption).

Generally speaking, considering that the reported errors are
at $1$-$\sigma$ and that the theoretical models are COBE normalized,
which allows a further $10 \%$ shift in amplitude, the model-independent values are in very good agreement with those predicted by the $\Lambda$-CDM models. 
It is interesting to notice that for the full data set,
the subsequent peaks appear to be slightly lower in amplitude
than those expected in the concordance model with $n_S=1$. 
This favors a spectral index $n_S$ slightly lower than one.
However, there is a strong degeneracy
between $n_S$ and the optical depth $\tau_c$ 
(see e.g. \cite{stompor}) and models with $n_s=1$ can be
put in better agreement with the observations when increasing
$\tau_c$.

\begin{table*}
\caption{\label{fmaev} 
First 3 columns: 
1-$\sigma$ constraints on the parameters of the phenomenological model 
for three subsets of data: The full data set, LF data and HF data.
The allowed range for the same parameters in a database of 
COBE normalized theoretical models (no experimental data are considered) 
is also reported in the following $2$ columns for the case
of {\it weak} priors ($0.05<\Omega_{cdm}<0.5$, $0.15 <\Omega_bh^2 < 0.25$,
$0.55 < h < 0.88$, $0.80 < n_S < 1.10$) and {\it strong} priors
($0.10<\Omega_{cdm}<0.35$, $0.18 <\Omega_bh^2 < 0.22$,
$0.65 < h < 0.80$, $0.95 < n_S < 1.05$).
In the last column we also show the values for a COBE normalized 
{\it concordance} model with $\Omega_{cdm}=0.31$, 
$\Omega_{b}=0.04$, $h=0.7$ and $n_s=1$.
}
{\small
\begin{tabular}{|c|c|c|c|c|c|c|}
\hline
CMB & Full data set & High Frequency & Low Frequency & 
$\Lambda$-CDM & $\Lambda$-CDM & Concordance\\ 
 Observable& & Experiments & Experiments & & & \\
 & Phen. Fit & Phen. Fit & Phen. Fit & Weak Priors& Strong Priors& \\\hline\hline
$\Delta T_1$& $69.3_{-2.3}^{+2.4} \mu K$& $70.7_{-3.5}^{+4.8} \mu K$ & $68.9_{-4.1}^{+5.5} \mu K$ &$(48-110) \mu K$ & $(65-95) \mu K$&$74.9 \mu K$\\\hline 
$\Delta T_2$& $44.0_{-1.9}^{+1.7} \mu K$& $43.1_{-3.0}^{+3.6} \mu K$ & $46.8_{-4.3}^{+3.6} \mu K$ &$(32-76) \mu K$&$(46-66) \mu K$&$53.6 \mu K$\\\hline
$\Delta T_3$& $43.8_{-4.4}^{+2.1} \mu K$& $40.7_{-5.0}^{+4.3} \mu K$ & $54.4_{-11.2}^{+3.4} \mu K$ &$(36-67) \mu K$&$(49-60) \mu K$& $54.6 \mu K$\\\hline
$\Delta T_4$& $31.2_{-8.7}^{+5.7} \mu K$& $25.7_{-5.7}^{+3.1} \mu K$ & $27.9_{-7.3}^{+8.7} \mu K$ &$(22-40) \mu K$&$(26-37) \mu K$& $35.6 \mu K$\\\hline
$\Delta T_5$&  $19.8_{-3.8}^{+1.8} \mu K$ & $18.4_{-5.2}^{+3.2} \mu K$ & $20.6_{-17.3}^{+12.7} \mu K$ &$(19-36) \mu K$&$(22-32) \mu K$& $27.6 \mu K$\\\hline
$\ell_1$&$208.8_{-6.1}^{+6.2}$& $204.6_{-7.9}^{+11.4}$ & $206.8_{-22.0}^{+10.8}$ &$192-279$&$203-242$& $215$\\\hline
$\ell_2$&$550_{-45}^{+13}$& $505_{-21}^{+25}$ & $533_{-20}^{+25}$ &$464-696$&$488-594$&$514$\\\hline
$\ell_3$&$824_{-41}^{+12}$& $764_{-42}^{+74}$ & $806_{-36}^{+26}$ &$692-1086$&$732-919$&$781$\\\hline
$\ell_4$& $1145_{-45}^{+30}$ & $1158_{-67}^{+242}$ & $1189_{-87}^{+32}$ &$1140-1386$&$1210-1301$&$1190$\\\hline
$\ell_5$& $1474_{-79}^{+153}$ & $1649_{-262}^{+142}$ & $1515_{-346}^{+81}$ &$1380-1590$&$1460-1550$&$1491$\\\hline
$\sigma_1$&$93.3_{-5.2}^{+4.5}$& $90.3_{-6.2}^{+8.2}$ & $88.2_{-12.3}^{+12.7}$ &$86-136$&$86-108$&$93$\\\hline
$\sigma_2$&$111.2_{\mbox{\tiny unbounded}}^{+27.7} $& $78.2_{-12.2}^{+19.3}$&$61.9_{\mbox{\tiny unbounded}}^{+36.7}$ & $71-121$&$71-107$&$86$\\\hline
$\sigma_3$&$82.5_{-23.1}^{+20.7}$& $136.3_{\mbox{\tiny unbounded}}^{+94.2}$ & $69.8_{\mbox{\tiny unbounded}}^{+12.8}$ & $86-150$&$86-150$&$102$\\\hline
$\sigma_4$& not constrained & not constrained & not constrained & $70-120$&$70-110$&$87$\\\hline
$\sigma_5$& not constrained & not constrained & not constrained &$70-120$&$75-110$&$88$\\\hline
$\ell_1/\ell_2$&$0.382_{-0.024}^{+0.033}$& $0.415_{\mbox{\tiny unbounded}}^{+0.014} $& $0.376_{-0.042}^{+0.021} $&$0.379-0.429$&$0.402-0.422$&$0.418$\\\hline
$\ell_1/\ell_3$&$0.256_{-0.016}^{+0.010}$& $0.266_{-0.040}^{+0.012}$ & $0.251_{-0.013}^{+0.012}$ &$0.242-0.290$&$0.259-0.281$&$0.275$\\\hline
$\Delta T_1/\Delta T_2$&$1.56_{-0.07}^{+0.04}$& $1.60_{-0.12}^{+0.17}$ & $1.44_{-0.15}^{+0.15}$ &$1.23-1.68$&$1.32-1.51$&$1.40$\\\hline
$\Delta T_1/\Delta T_3$&$1.62_{-0.11}^{+0.06}$& $1.65_{-0.10}^{+0.15} $& $1.29_{-0.18}^{+0.22} $&$1.12-2.13$&$1.26-1.73$&1.37\\\hline
$\Delta T_2/\Delta T_3$&$1.01_{-0.09}^{+0.10}$& $0.99_{-0.12}^{+0.20} $& $0.87_{-0.10}^{+0.13} $&$0.80-1.32$&$0.92-1.17$&$0.98$\\\hline
$\ell_2-\ell_1$&$332_{-42}^{+19}$& $290_{-19}^{+39}$ & $329_{-30}^{+47}$ &$267-427$&$284-353$&$299$\\\hline
$\ell_3-\ell_1$&$611_{-39}^{+26}$& $555_{-41}^{+97}$ & $605_{-30}^{+31}$ &$484-817$&$528-679$&$566$\\\hline
\hline
\end{tabular}
}
\end{table*}

\begin{table*}
\caption{\label{corre} 
Correlation matrix for the $15$ phenomenological parameters when fitted to the full data set.}
{\small
\begin{tabular}{|c||c|c|c|c|c|c|c|c|c|c|c|c|c|c|c|}
\hline
&$\Delta T_1$&	$\Delta T_2$& $\Delta T_3$& $\Delta T_4$& $\Delta T_5$&
$\ell_1$ & $\ell_2$& $\ell_3$& $\ell_4$& $\ell_5$
&$\sigma_1$&$\sigma_2$&$\sigma_3$&$\sigma_4$&$\sigma_5$\\\hline\hline	
$\Delta T_1$&$1.00$&$0.20$&$0.32$&$0.12$&$0.06$&$-0.02$&$-0.17$&$-0.07$&$0.03$&$-0.02$&$-0.23$&$-0.19$&$0.12$&$-0.18$&$0.04$\\\hline
$\Delta T_2$&$0.20$&$1.00$&$0.03$&$-0.06$&$-0.07$&$-0.30$&$0.54$&$0.54$&$-0.15$&$0.00$&$-0.32$&$0.38$&$-0.70$&$0.18$&$0.11$\\\hline
$\Delta T_3$&$0.32$&$0.03$&$1.00$&$0.37$&$0.07$&$0.38$&$-0.54$&$-0.27$&$0.40$&$0.08$&$0.30$&$-0.63$&$0.40$&$-0.45$&$-0.08$\\\hline
$\Delta T_4$&$0.12$&$-0.06$&$0.37$&$1.00$&$0.32$&$0.06$&$-0.19$&$0.08$&$0.20$&$-0.10$&$0.03$&$-0.13$&$0.27$&$-0.71$&$-0.10$\\\hline
$\Delta T_5$&$0.06$&$-0.07$&$0.07$&$0.32$&$1.00$&$-0.03$&$0.01$&$0.16$&$-0.36$&$-0.61$&$-0.07$&$0.05$&$0.06$&$-0.51$&$-0.09$\\\hline
$\ell_1$&$-0.02$&$-0.30$&$0.38$&$0.06$&$-0.03$&$1.00$&$-0.42$&$-0.55$&$0.10$&$0.08$&$0.60$&$-0.75$&$0.49$&$-0.09$&$-0.05$\\\hline
$\ell_2$&$-0.17$&$0.54$&$-0.54$&$-0.19$&$0.01$&$-0.42$&$1.00$&$0.72$&$-0.37$&$-0.09$&$-0.32$&$0.79$&$-0.84$&$0.31$&$0.08$\\\hline
$\ell_3$&$-0.07$&$0.54$&$-0.27$&$0.08$&$0.16$&$-0.55$&$0.72$&$1.00$&$-0.04$&$-0.13$&$-0.52$&$0.77$&$-0.58$&$-0.07$&$0.01$\\\hline
$\ell_4$&$0.03$&$-0.15$&$0.40$&$0.20$&$-0.36$&$0.10$&$-0.37$&$-0.04$&$1.00$&$0.64$&$0.10$&$-0.23$&$0.48$&$-0.20$&$-0.25$\\\hline
$\ell_5$&$-0.02$&$0.00$&$0.08$&$-0.10$&$-0.61$&$0.08$&$-0.09$&$-0.13$&$0.64$&$1.00$&$0.11$&$-0.09$&$0.10$&$0.34$&$-0.19$\\\hline
$\sigma_1$&$-0.23$&$-0.32$&$0.30$&$0.03$&$-0.07$&$0.60$&$-0.32$&$-0.52$&$0.10$&$0.11$&$1.00$&$-0.66$&$0.42$&$-0.01$&$-0.06$\\\hline
$\sigma_2$&$-0.19$&$0.38$&$-0.63$&$-0.13$&$0.05$&$-0.75$&$0.79$&$0.77$&$-0.23$&$-0.09$&$-0.66$&$1.00$&$-0.74$&$0.20$&$0.05$\\\hline
$\sigma_3$&$0.12$&$-0.70$&$0.40$&$0.27$&$0.06$&$0.49$&$-0.84$&$-0.58$&$0.48$&$0.10$&$0.42$&$-0.74$&$1.00$&$-0.46$&$-0.09$\\\hline
$\sigma_4$&$-0.18$&$0.18$&$-0.45$&$-0.71$&$-0.51$&$-0.09$&$0.31$&$-0.07$&$-0.20$&$0.34$&$-0.01$&$0.20$&$-0.46$&$1.00$&$-0.13$\\\hline
$\sigma_5$&$0.04$&$0.11$&$-0.08$&$-0.10$&$-0.09$&$-0.05$&$0.08$&$0.01$&$-0.25$&$-0.19$&$-0.06$&$0.05$&$-0.09$&$-0.13$&$1.00$\\\hline
\hline
\end{tabular}
}
\end{table*}

\medskip

\medskip

\medskip

\begin{figure}[ht]
\centerline{\psfig{figure=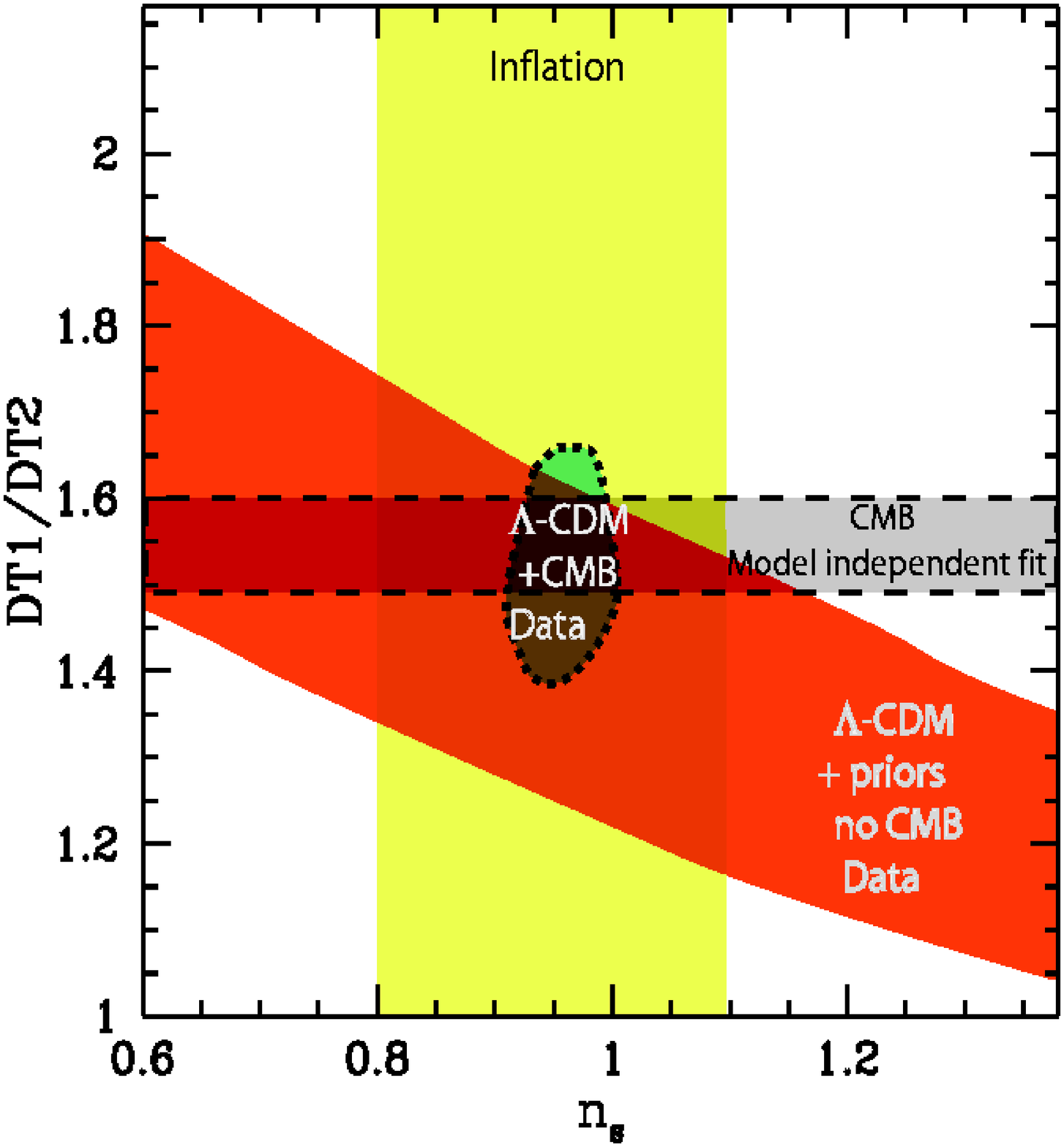,width=5.3cm}}
\centerline{\psfig{figure=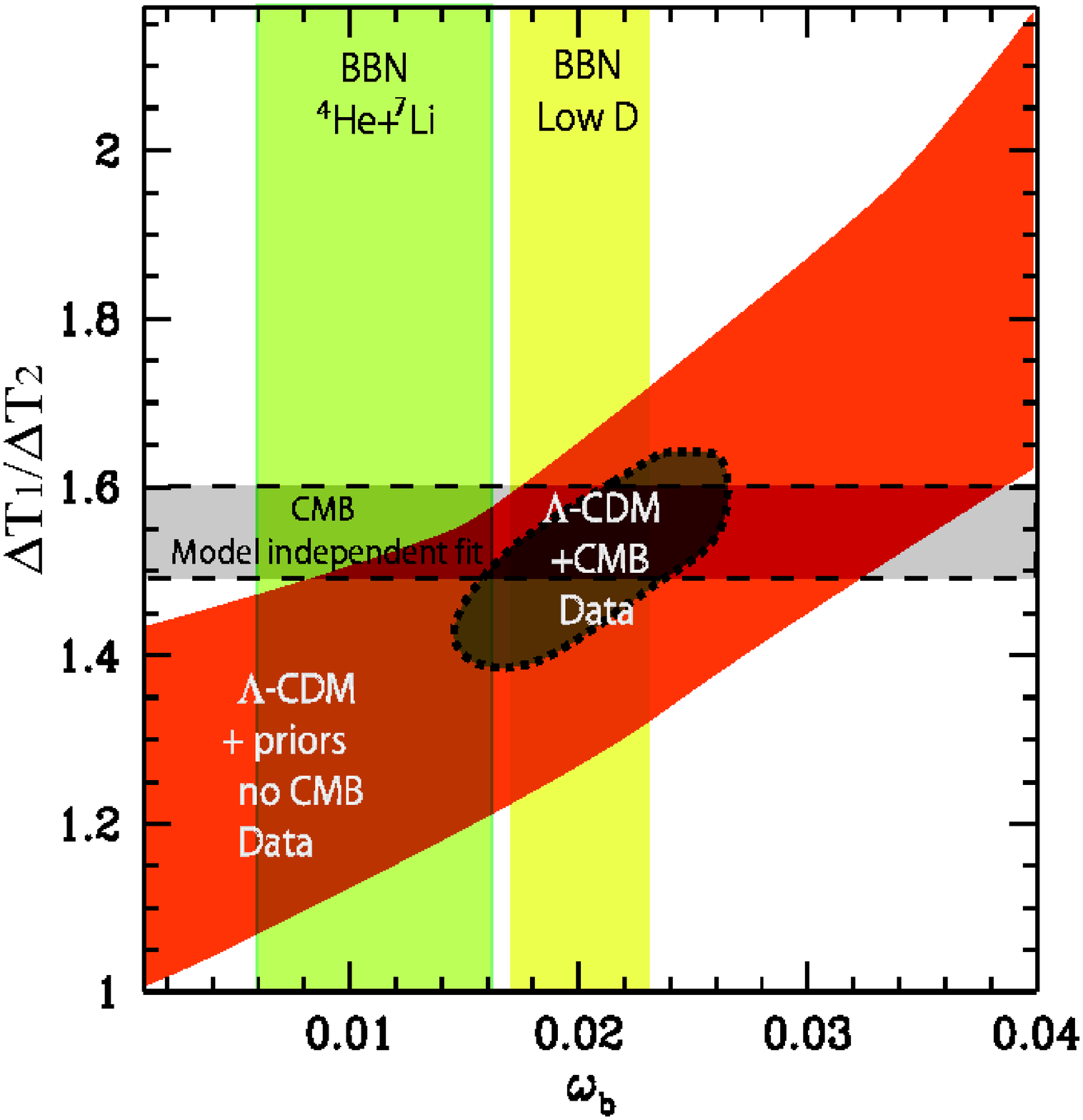,width=5.3cm}}
\caption{Observed relative amplitude between the
first and second peak and comparison with adiabatic 
$\Lambda$-CDM cosmological models.
The red region (``$\Lambda$-CDM no CMB'') defines
all the values of $\Delta T_1/\Delta T_2$ present in the model
template described in the text with the additional 
priors $0.1<\Omega_{cdm}<0.5$ and
$0.55<h<0.88$. In the top panel, the additional
prior $0.015 < \Omega_bh^2 <0.025$ is included, while in
the bottom panel we use $0.8 < n_S < 1.1$.
The $68 \%$ c.l. 
constraint obtained by the phenomenological fit and 
he $95 \%$ c.l. of all the $\Lambda-CDM$ models
compatible with CMB are also plotted for comparison.
The constraints on $\Omega_bh^2$ from standard BBN 
from $D$ (Burles et al.2001) and $^4He+^7Li$ (Cyburt et al. 2001)
observations are also included in the second plot. }

\label{dt1dt2nsob}
\end{figure}

\medskip

\begin{figure}[ht]
\centerline{\psfig{figure=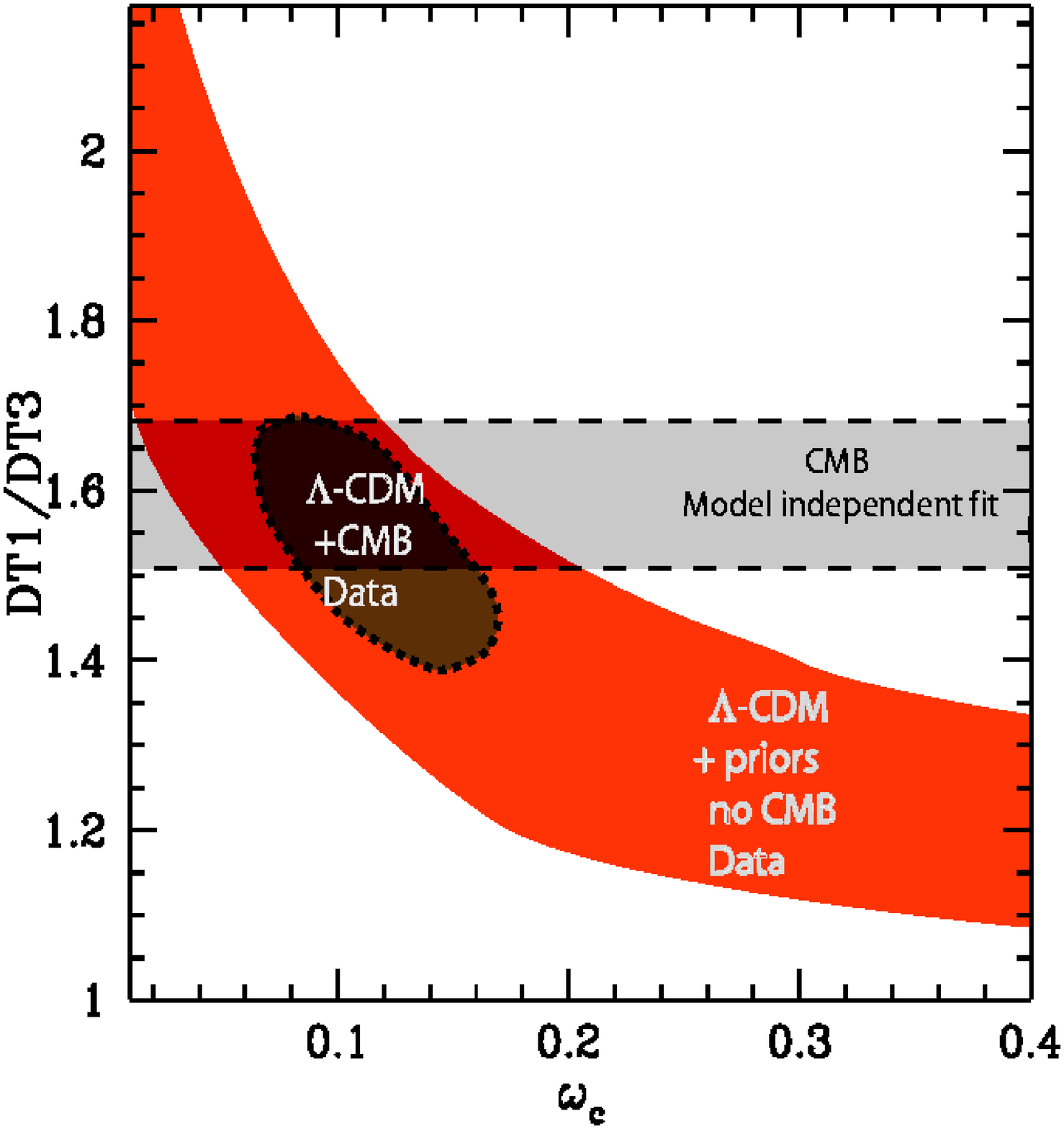,width=5.3cm}}
\centerline{\psfig{figure=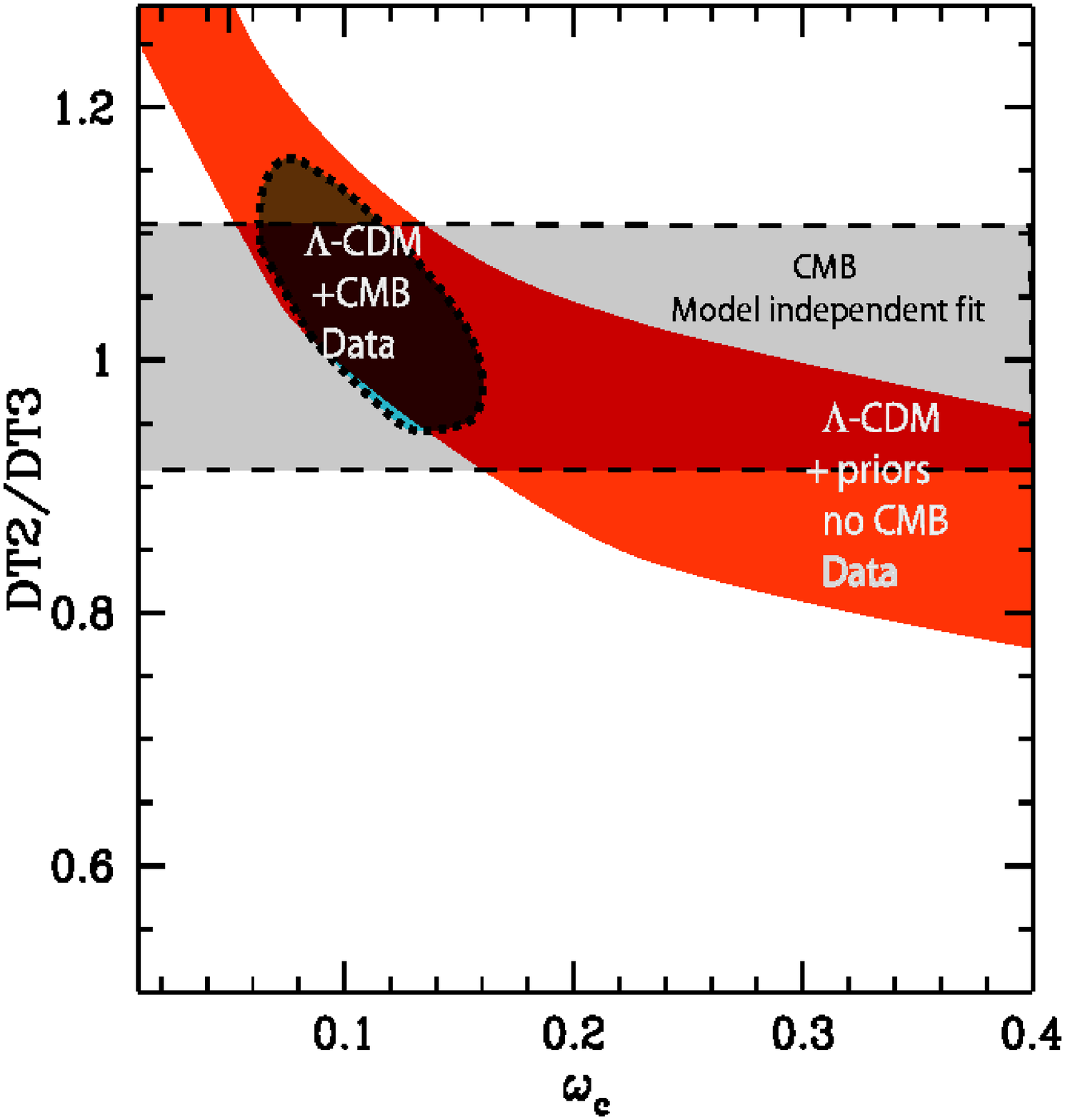,width=5.3cm}}
\caption{Observed relative amplitudes between the first and third peaks, and between the second and third peaks, and comparison with adiabatic $\Lambda$-CDM cosmological models. The red region (``$\Lambda$-CDM no CMB'') defines all the values of $\Delta T_1/\Delta T_3$ present in the template of models described in the text with the additional priors $0.55<h<0.88$, $0.015 < \Omega_bh^2 <0.025$ and $0.8 < n_S < 1.1$. The $68 \%$ c.l. constraint obtained by the phenomenological fit and the $95 \%$ c.l. of all the $\Lambda$-CDM models compatible with CMB are also plotted for comparison.}\label{dt1dt2dt3oc}\end{figure}\medskip

\begin{figure}[ht]
\centerline{\psfig{figure=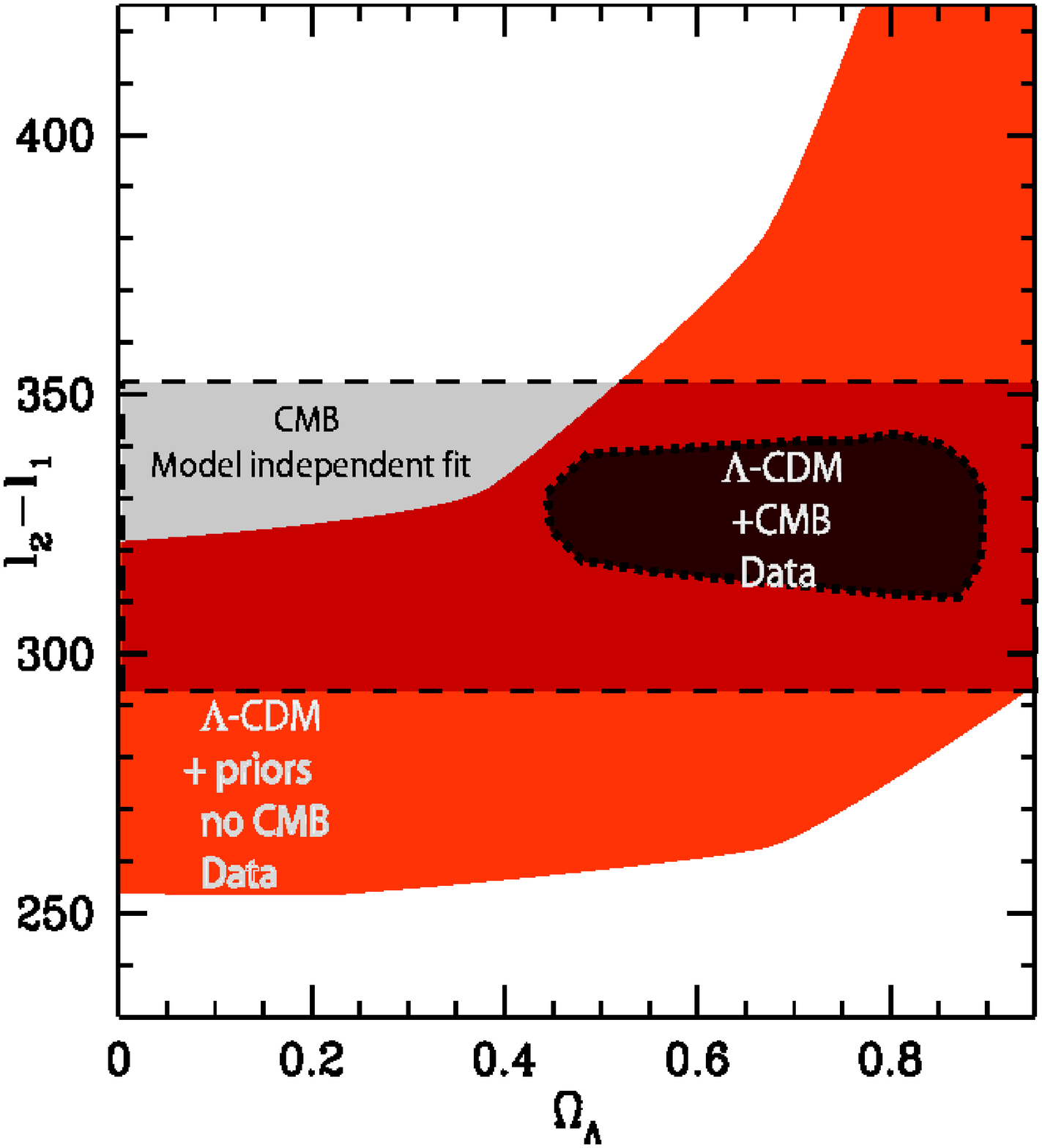,width=5.3cm}}
\centerline{\psfig{figure=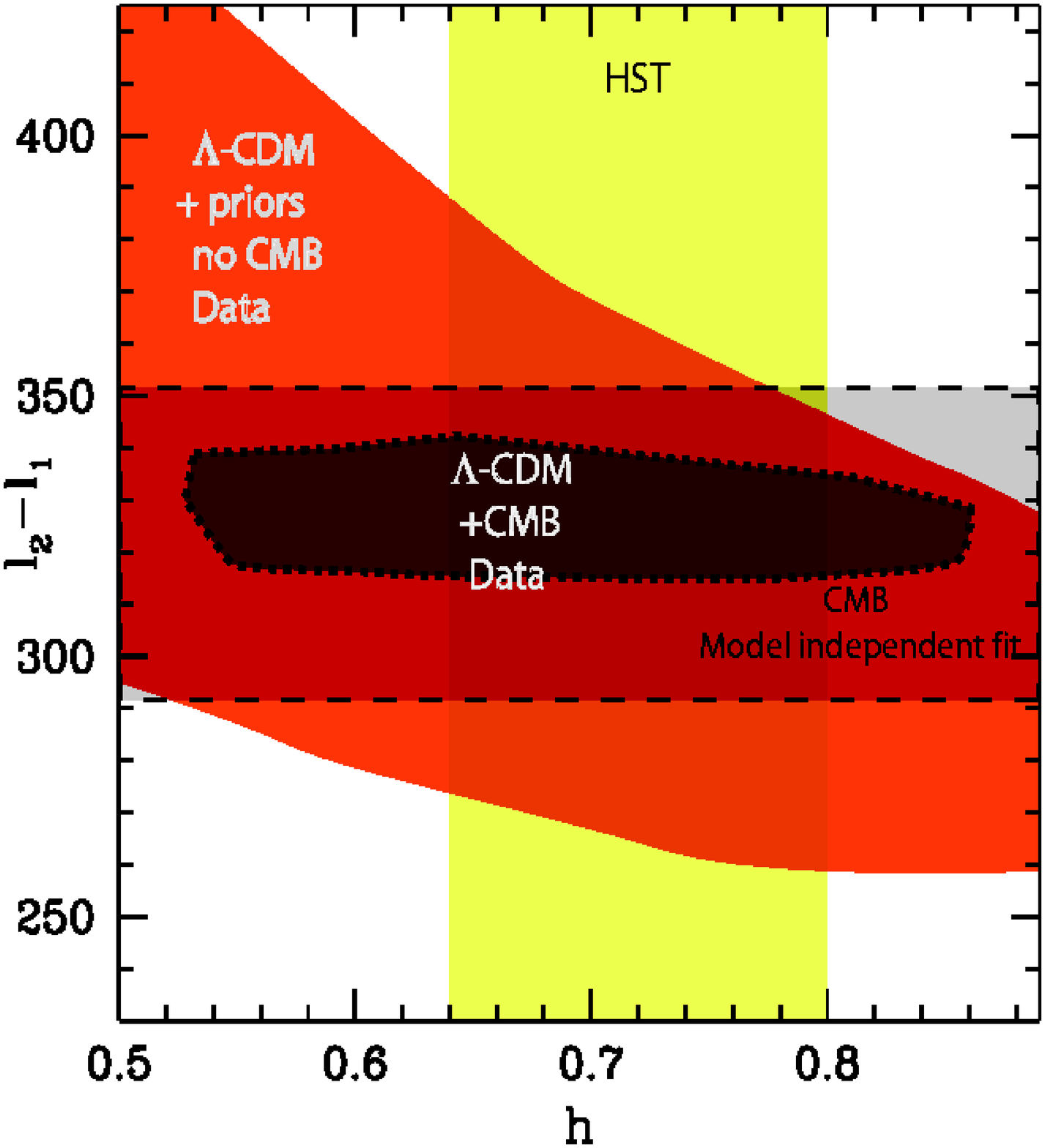,width=5.3cm}}
\centerline{\psfig{figure=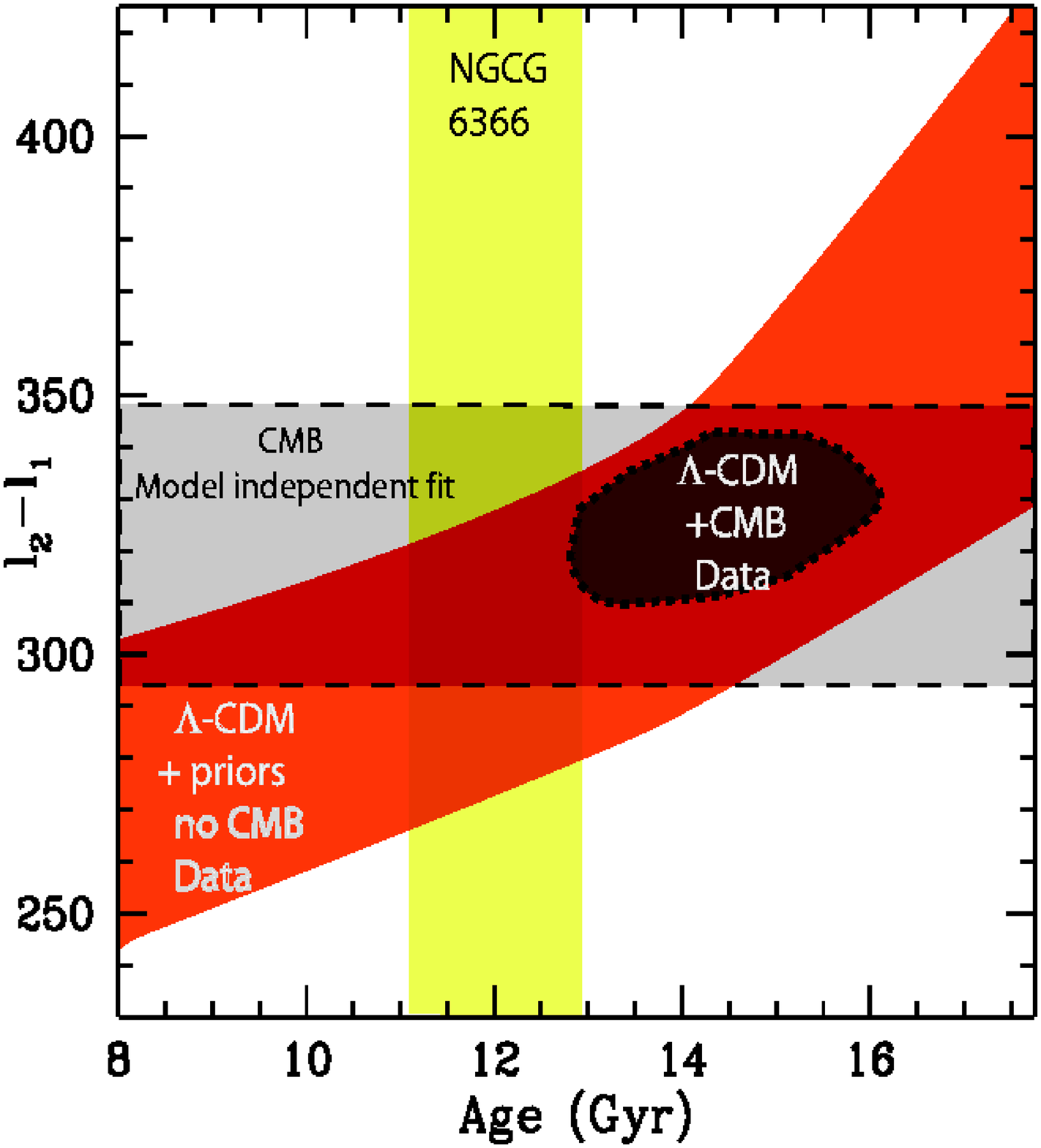,width=5.3cm}}
\caption{Observed relative positions between the
first and second peak and comparison with adiabatic 
$\Lambda$-CDM cosmological models.
The red region (``$\Lambda$-CDM no CMB'') defines
all the values of $\ell_2-\ell_1$ present in the model
template described in the text with the additional 
priors $0.015 < \Omega_bh^2 <0.025$ and 
$0.8 < n_S < 1.1$. 
In the top panel (constraints on $\Omega_{\Lambda}$)
the additional prior $0.55<h<0.88$ is used.
In the centre panel, we use $0.1<\Omega_{cdm}<0.5$.
The band on the $y$-axis on the bottom panel is the
constraint on the age of the oldest halo globular cluster 
in the sample of Salaris and Weiss ($1998$).
The $68 \%$ c.l. constraint obtained by the phenomenological fit and 
the $95 \%$ c.l. of all the $\Lambda$-CDM models
compatible with CMB are also plotted for comparison.}
\label{l2l1age}
\end{figure}

\begin{figure}[ht]
\centerline{\psfig{figure=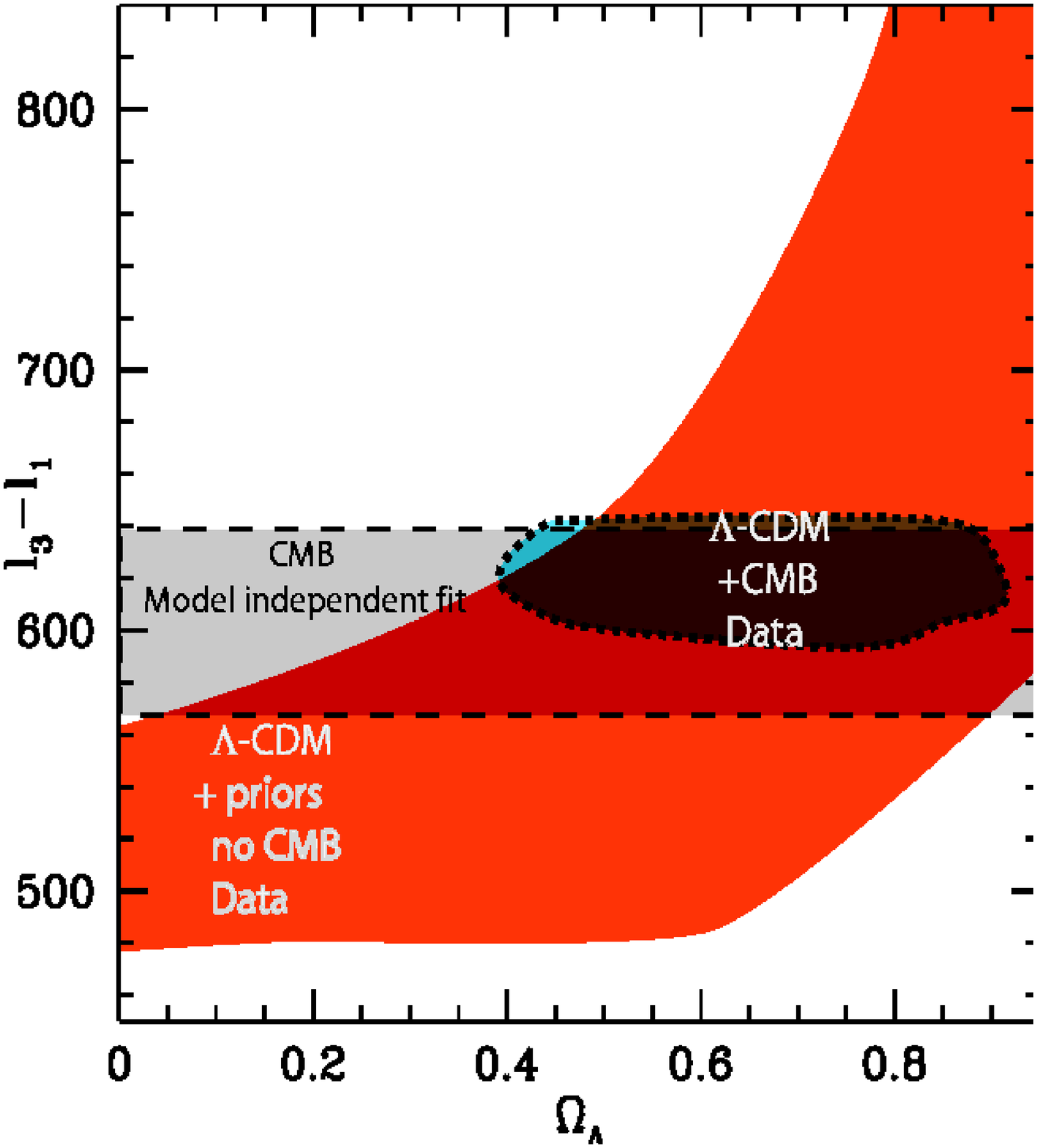,width=5.3cm}}
\centerline{\psfig{figure=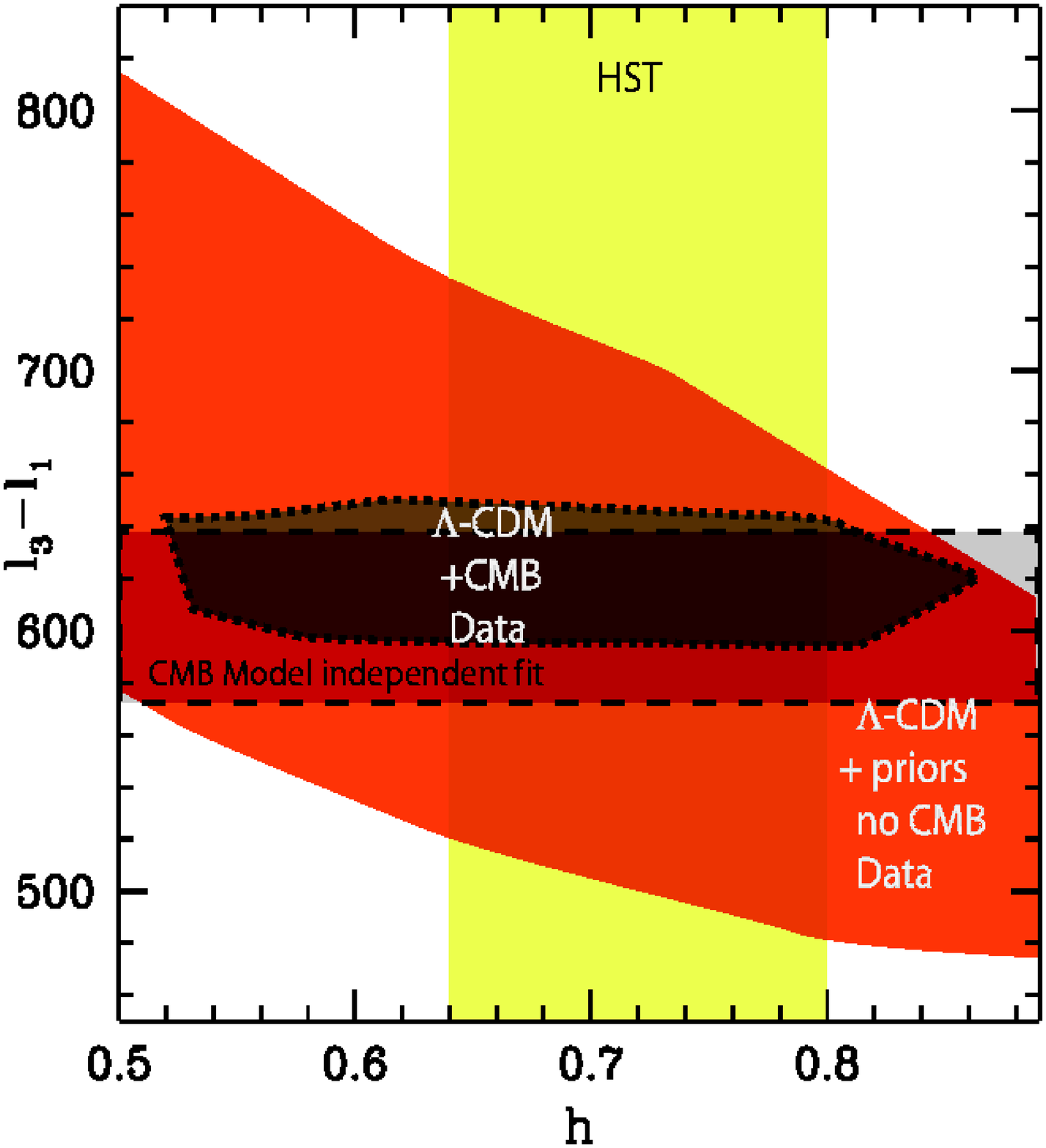,width=5.3cm}}
\centerline{\psfig{figure=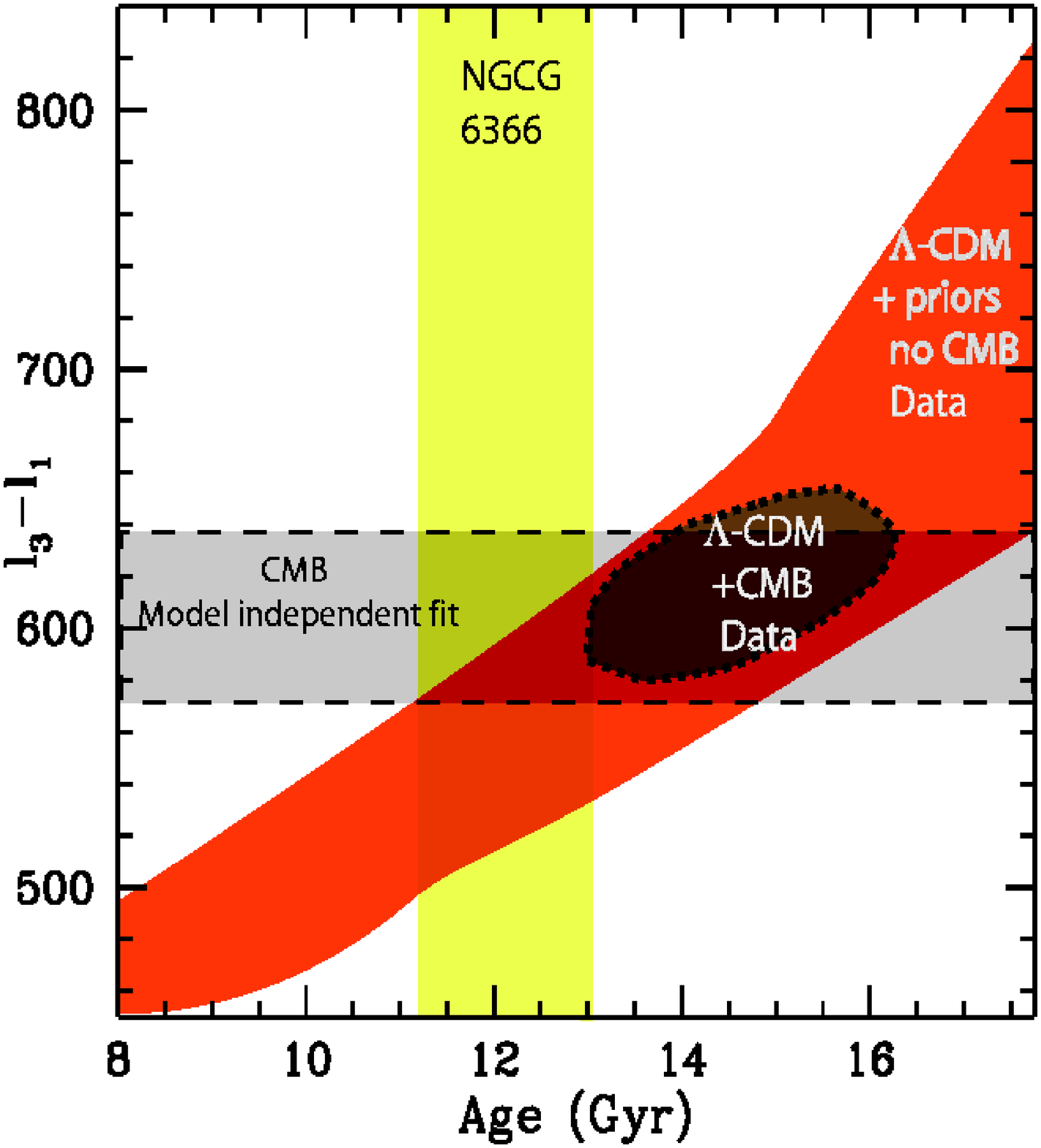,width=5.3cm}}
\caption{Observed relative positions between the
first and third peak and comparison with adiabatic 
$\Lambda$-CDM cosmological models. The definitions of
the regions and priors used are the same as in Fig. \ref{l2l1age}}
\label{l3l1age}
\end{figure}

\begin{figure}[ht]
\centerline{\psfig{figure=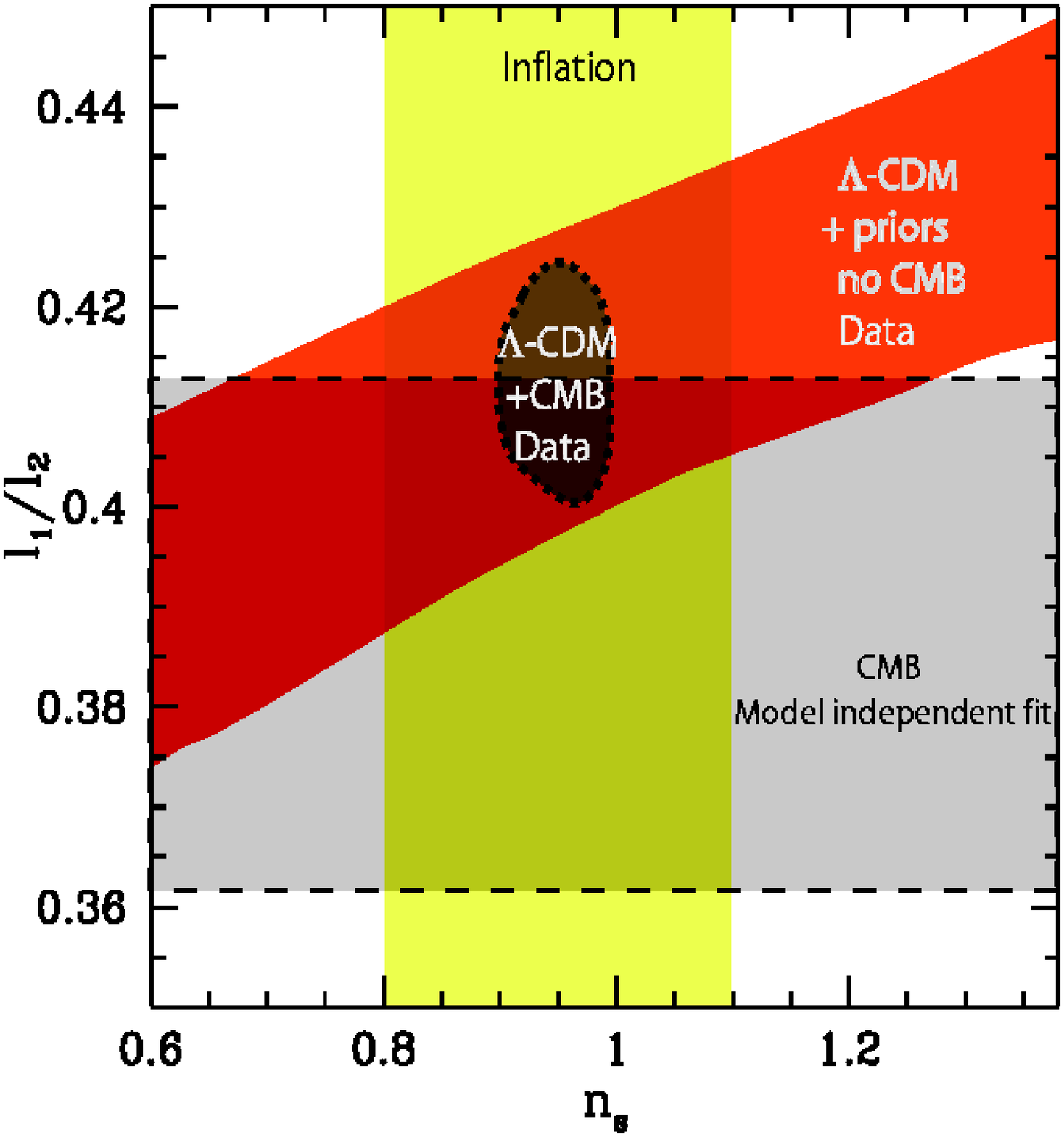,width=5.3cm}}
\centerline{\psfig{figure=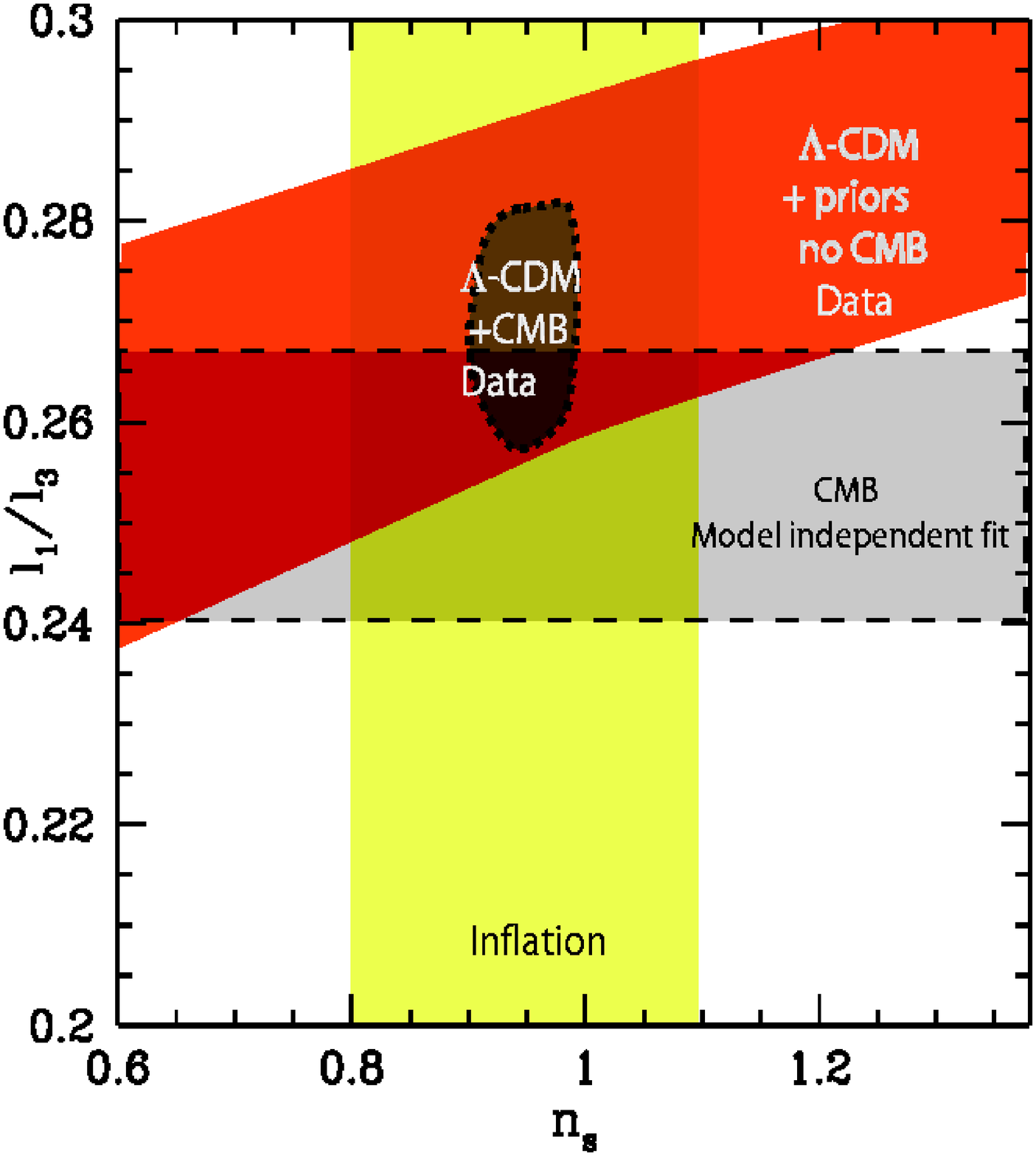,width=5.3cm}}
\caption{Observed relative positions between the
first, second and third peak and comparison with adiabatic 
$\Lambda$-CDM cosmological models.The red region (``$\Lambda$-CDM no CMB'') defines all the values of $\ell_1/\ell_2$ and $\ell_1/\ell_3$ present in the modeltemplate described in the text with the additional priors $0.015 < \Omega_bh^2 <0.025$, $0.55<h<0.88$ and $0.1<\Omega_{cdm}<0.5$.The $68 \%$ c.l. constraint obtained by the phenomenological fit and the $95 \%$ c.l. of all the $\Lambda-CDM$ models compatible with CMB are also plotted for comparison.}\label{l1l2l3ns}
\end{figure}

We further investigate the consistency with 
$\Lambda$-CDM by considering phenomenological diagrams 
relating the relative amplitudes and positions 
of the peaks with variations in a specific 
physical parameter. 

In the $\Lambda$-CDM adiabatic scenario, two key parameters
control the relative power between the first and second
peaks: the physical baryon density $\omega_b$ and the
primordial spectral index $n_S$ (see e.g. \cite{bibbia}).
Increasing $\omega_b$ enhances the odd-numbered peaks 
relative to the even-numbered ones, while increasing $n_S$ enhances
the small-scale peaks relative to the ones at larger scales.

In figure \ref{dt1dt2nsob} we plot the values
allowed in our model template (restricted by a set of 
rather conservative cosmological constraints, see the caption) 
for the relative amplitude $\Delta T_1/\Delta T_2$ 
as functions of the parameters $\omega_b$ and $n_S$.
As expected, increasing (decreasing) $\omega_b$ ($n_S$) 
increases $\Delta T_1/\Delta T_2$.
The region is very broad, mainly owing to the degeneracy between these two parameters. Nevertheless, super-imposing the 1-$\sigma$
phenomenological constraint on $\Delta T_1/\Delta T_2$ in the diagram
provides interesting constraints.
Models with a value of the spectral index $n_s > 1.15$, which are
not easily accommodated in most inflationary models 
(see e.g. \cite{kinney} and references therein) or in disagreement with the 
BBN constraint $\omega_b=0.020\pm0.002$ 
are in fact not favored by our model-independent fit. In that figure, we also plot the constraints obtained by fitting the CMB data with the models in the template. This reduces in a severe way the number of allowed $\Lambda$-CDM models. Nevertheless, the result on relative amplitude of the peaks is completely consistent with the one derived by the phenomenological fit. This method provides better constraints on the amount of
cold dark matter $\omega_{cdm}$ if we consider the relative 
amplitudes $\Delta T_1/\Delta T_3$ and $\Delta T_2/\Delta T_3$ as we do in
the top and bottom panels of figure ~\ref{dt1dt2dt3oc}.
A decrease in $\omega_{cdm}$ has the effect of decreasing
the amplitude of the third peak (see e.g. \cite{gms}).
As we can see, the two observational values of
$\Delta T_1/\Delta T_3$ and $\Delta T_2/\Delta T_3$ provide 
similar constraints on $\omega_{cdm}$ in a non trivial way
with $\omega_{cdm}\sim 0.12$
and with $\omega_{cdm}=0$ or $\omega_{cdm}>0.3$ in disagreement 
with the data.

In a flat $\Lambda$-CDM model a variation in $\Omega_{\Lambda}$
shifts the spectrum as $\ell \rightarrow {\cal R} \ell$ with
the shift parameter ${\cal R}$ given by
(see \cite{efsbond}, \cite{melou}):

\begin{equation}
{\cal R} =\sqrt{|\Omega_m|}\int_0^{z_{dec}}
{[(1-\Omega_{\Lambda})(1+z)^3+\Omega_{\Lambda}]^{-1/2} dz}.
\end{equation}

Varying the Hubble constant, parametrized as $H_0=100 h$ Km/sec/Mpc, 
changes the scale of equality and produces a similar shift.
These two parameters are related to  the age of the universe by:

\begin{equation}
t_0= - 9.8 Gy \int_{\inf}^0{{ 
{[h^2((1-\Omega_{\Lambda})(1+z)^3+\Omega_{\Lambda})]^{-1/2}}} dz}
\end{equation}

We can therefore expect that a determination of the
peak positions provides constraints on these three quantities.

In figure \ref{l2l1age} and figure \ref{l3l1age} we plot similar diagrams as
above for $\ell_2-\ell_1$ and $\ell_3-\ell_1$ as functions
of $\Omega_{\Lambda}$, $h$ and age, $t_0$.
Even if the region of the allowed models is
quite broad because of the intrinsic degeneracies, the observed
peak positions strongly favor a model with
cosmological constant $\Omega_{\Lambda}> 0$, 
a Hubble parameter $h<0.8$, compatible with the
Hubble Space Telescope (HST) result of $h=0.72 \pm 0.08$ 
(\cite{freedman}), and an age $t_0> 14$ Gy, compatible with the
age of the oldest globular clusters (see e.g. \cite{age}).
Again, the values obtained by the phenomenological fit
are in agreement with those derived by the standard
CMB+$\Lambda$-CDM analysis. 

Another parameter that affects the position of
the peaks is the spectral index $n_S$.
However, the effect is different, the shift being  
scale dependent. Therefore, it is better to consider 
the quantities $\ell_1/\ell_2$ and $\ell_1/\ell_3$ which
are unaffected by the overall shift $\cal R$.
The corresponding diagrams are plotted in figure \ref{l1l2l3ns}.
As mentioned earlier, the observed values point towards a low value of the spectral index $n_S \le 1$.

\medskip
\section{Conclusions}
\medskip

In this {\it paper} we investigated the consistency 
of the most recent CMB data with a class
of $\Lambda$-CDM adiabatic inflationary models.
First we characterized the positions, amplitudes and widths
of the peaks by fitting the data with simple phenomenological
functions composed by several gaussians.
The detection of the peak amplitudes and positions is quite 
robust and stable between different data sets.
We found that all the features are consistent with 
those expected by the standard theory.
We also examined where the data contains the most information in the
power-angular scale plane. We found that the low frequency experiments provide
good constraints at small angular scales, consistent with the expected 
damping tail, whereas high frequency experiments provide strong limits on 
the power at large and intermediate scales. We observe that HF experiments 
and LF experiments yield very consistent results, although LF data seems 
to provide evidence for higher secondary oscillations.
Overall, the power spectrum is now well determined
until $\ell \sim 1500$. The inclusion of older data does not affect our
conclusions as they do not measure the power beyond $\ell \sim 400$.

Furthermore, we related the features in the spectrum
with several cosmological parameters by introducing 
cosmological diagrams that can be used for quick, by-eye, 
parameter estimations.

The relative amplitude of the first and second peak, in
particular, of about $\sim 1.56$ 
is consistent with the baryon density expected from BBN 
and suggests a value of $n_S$ lower than one in the case of negligible
reionization.
The amplitudes of the third peak relative to the first and to
the second, $\Delta T_1/\Delta T_3\sim 1.6$ and 
$\Delta T_2/\Delta T_3 \sim 1$ strongly suggest the presence of 
cold dark matter but also limits time its contribution to values  
$\omega_{cdm} <0.2$.
The relative positions of the peaks, $\ell_2-\ell_1 \sim 330$ and
$\ell_3-\ell_1 \sim 610$ is pointing towards the
presence of a cosmological constant, a Hubble parameter 
on the low side of the value allowed by the recent HST measurements
($h \sim 0.65$) and to an age of the universe
$t_0 \sim 14.5$ Gyrs consistent with the measurements of the oldest
globular clusters.

It is reassuring that all those conclusions, obtained
by just drawing few lines in the diagrams presented in Figs. $5-9$,
are in agreement with the results obtained by a more careful 
standard analysis. 
Within the models considered in our database we found 
(at $68 \%$ c.l.): $n_s=0.96\pm 0.03$, $\omega_b=0.022\pm 0.003$,
$\omega_{cdm}=0.12 \pm 0.03$, $\Omega_{\Lambda}=0.63\pm0.16$, and 
$t_0= 14.2 \pm 0.7$ Gyrs.

The results obtained here show no need for modifications to the 
standard model, like gravity waves, quintessence, 
isocurvature modes, or 
extra-backgrounds of relativistic particles.
Furthermore, possible systematic effects due
to unknown foregrounds or calibration and beam 
uncertainties are not immediately suggested, since
the different data sets are consistent with the theory.

Even if the width of the gaussians is poorly constrained, we found 
supporting evidence for multiple oscillations in 
the data between $430 < \ell < 910$. Beyond that, the newest 
experimental results show a damping of the power.

It is the duty of future satellite CMB experiments 
to point out discrepancies that might place the possibility 
of new physics in a more favorable light.

\medskip 

\textit{Acknowledgements} 
It is a pleasure to thank Ruth Durrer, Anthony Lewis, Ruediger Kneissl, 
Roya Mohayaee, Lyman Page, Joseph Silk and Anze Slosar for useful 
comments. We acknowledge the use of CMBFAST~\cite{sz}.
CJO is supported by the Leenaards Foundation, the Acube Fund, 
an Isaac Newton Studentship and a Girton College Scholarship.
AM is supported by PPARC.

\end{document}